\documentclass[aps, reprint, prl, superscriptaddress]{revtex4-1}

\usepackage{amsmath,amsthm,amsfonts,bm}
\usepackage{graphicx,xcolor}

\newcommand{\bmh}[1]{\bm{\hat #1}}
\newcommand{\sgn}{\mathrm{sgn}}
\newcommand{\ud}{\updownarrow}

\begin{document}
\title{Complete absorption of topologically protected waves}
\author{Guido Baardink}
\affiliation{Department of Physics, University of Bath, Claverton Down, Bath, BA2 7AY, UK}
\author{Gino Cassella}
\affiliation{Department of Physics, University of Bath, Claverton Down, Bath, BA2 7AY, UK}
\author{Luke Neville}
\affiliation{Department of Physics, University of Bath, Claverton Down, Bath, BA2 7AY, UK}
\author{Paul A.~Milewski}
\affiliation{Department of Mathematical Sciences, University of Bath, Claverton Down, Bath, BA2 7AY, UK}
\author{Anton Souslov}
\email{a.souslov@bath.ac.uk}
\affiliation{Department of Physics, University of Bath, Claverton Down, Bath, BA2 7AY, UK}
\date{\today}

\begin{abstract}
Chiral edge states can transmit energy along imperfect interfaces in a topologically robust and unidirectional manner when protected by bulk-boundary correspondence.
However, in continuum systems, the number of states at an interface can depend on boundary conditions.
Here we design interfaces that host a net flux of the number of modes into a region, trapping incoming energy.
As a realization, we present a model system of two topological fluids composed of counter-spinning particles, which are separated by a boundary that transitions from a fluid-fluid interface into a no-slip wall.
In these fluids, chiral edge states disappear, which implies non-Hermiticity and leads to a novel interplay between topology and energy dissipation.
Solving the fluid equations of motion, we find explicit expressions for the disappearing modes. 
We then conclude that energy dissipation is sped up by mode trapping.
Instead of making efficient waveguides, our work shows how topology can be exploited for applications towards acoustic absorption, shielding, and soundproofing.
\end{abstract}

\maketitle

\section{I. Introduction}
When does broken time-reversal symmetry imply non-Hermiticity? The converse is always true, because gains and losses in non-Hermitian Hamiltonians break thermodynamic reversibility and time-reversal symmetry{~\cite{brandenbourger2019non,ghatak2020observation,Rosa2020,helbig2020generalized,scheibner2020odd,schomerus2020nonreciprocal,Budich2019,gong2018topological,yoshida2019exceptional,shen_topological_2018,alvarez2018non,borgnia2020non,lee2019topological}}. 
By contrast, transverse velocity-dependent forces, although odd under time reversal, do conserve energy and preserve Hermiticity. 
Examples of these Hermitian effects include Lorentz forces due to an applied magnetic field~\cite{hasan2010colloquium}, Coriolis forces due to an external rotation~\cite{Kariyado2015,Shankar2017,delplace2017topological}, and so-called odd (Hall) viscosity{~\cite{Avron1995,Avron1998,banerjee2017odd,soni2019odd,souslov2019topological}}
arising in exotic quantum and classical fluids.
In gapped Hermitian systems, broken time-reversal symmetry is associated with a wealth of topological phenomena such as unidirectional interface states protected against scattering.
Such chiral interface modes are guaranteed to accompany any jump in a topological invariant called the Chern number due to bulk-interface correspondence~\cite{hasan2010colloquium,Shankar2020}. 
These robust edge states have been proposed and realized across a variety of electronic~\cite{von_klitzing_quantized_1986,thouless_quantized_1982,haldane_model_1988,hasan2010colloquium}, photonic{~\cite{rechtsman_photonic_2013,lu_topological_2014,ozawa2019topological}}, and mechanical{~\cite{Nash2015,wang_topological_2015,Wang2015a,Kariyado2015,Shankar2017,Dasbiswas2017,mitchell_amorphous_2018,souslov2019topological,susstrunk_classification_2016,bertoldi2017flexible,Shankar2020}} systems.

\begin{figure}[t!]
    \centering
    \includegraphics[width=\linewidth]{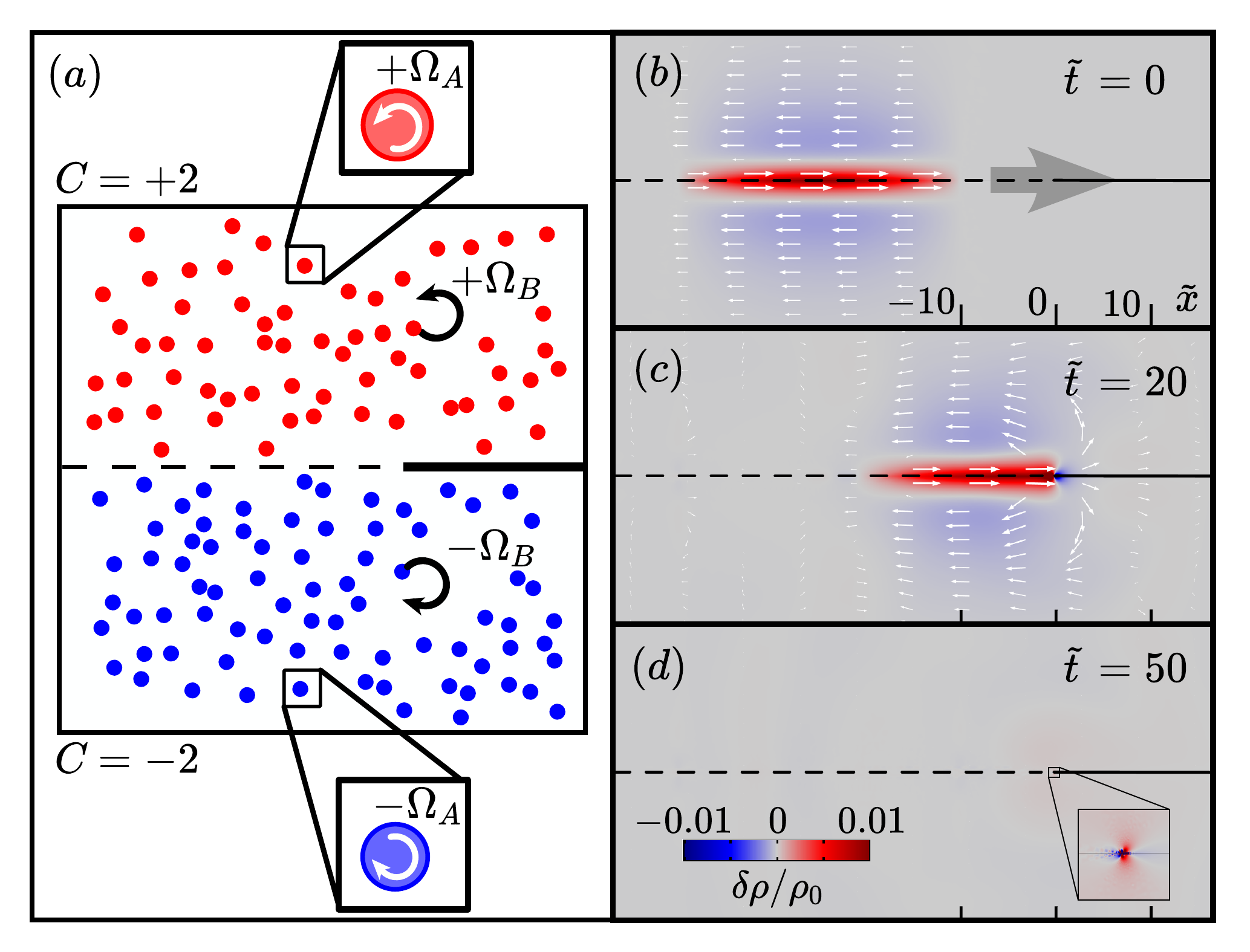}
    \caption{
    Completely absorbed interface mode in a chiral active fluid. 
    (a) Schematic a two-dimensional system of two chiral active fluids composed of counter-rotating particles. 
    Each particle spins with rotational speed $\pm\Omega_A$ and is subject to a Coriolis force density $\pm \Omega_B \bm v^*$, where $\bm v^*$ is the velocity $\bm v$ rotated by $90^\circ$. 
    The bulk is gapped with topological invariant $C=\sgn(\Omega_A)+\sgn(\Omega_B)$.
    The interface between the two fluids supports right-moving edge modes, one of which is shown in (b)-(d). (b) The right-moving pulse along the fluid-fluid interface (dotted line). (c) The pulse hits the no-slip wall (solid line). (d) The pulse is completely absorbed.
    Density differences $\delta\rho$ around an equilibrium density $\rho_0$ are shown in color, and the velocity field is portrayed by white arrows. 
    Time $\tilde t = t/T$ and space $\tilde x= x/L$ are nondimensionalized by $L \equiv |\nu^o|/c$ and $T \equiv |\nu^o|/c^2$, using the odd viscosity parameter $\nu^o \sim \Omega_A$, see text. 
    The incoming wave has parameters $\Omega_B T=1/5$ and $\omega T = q L =\pi/30$.
    }\label{fig:1}
\end{figure}
 
In continuum { systems}, bulk-boundary correspondence requires additional ingredients{~\cite{Fukui2012,bal2019continuous,Silveirinha2016}}
and can be even be violated~{ \cite{Tauber2020,fernandes2019topological,gangaraj2020physical,massana2019edge}}. 
We consider the { two-dimensional system} illustrated in Fig.~\ref{fig:1}(a). The container includes an interface between two counter-rotating chiral active fluids composed of self-rotating particles (equivalently, active rotors), each spinning with an intrinsic rotation rate $\pm\Omega_A${~\cite{Tsai2005,Furthauer2012, Drescher2009, Petroff2015, Yan2015,  Snezhko2016, Maggi2015,  Spellings2015, vanZuiden2016}}.
In addition, either through external forces or internal active stresses, 
the fluid flow is characterized by vorticity $\pm\Omega_B$.
These fluids exhibit bulk acoustic band gaps, which in the presence of active rotation can be characterized by topological Chern number $C = \pm 2$~\cite{souslov2019topological}.
The dashed middle line denotes a fluid-fluid  (i.e., stress continuity) boundary, at which bulk-boundary correspondence holds~\cite{Shankar2017,Tauber2019}, resulting in 4 (four) interface modes. 
Along the right-most segment, we imagine a thin rigid slab is inserted and the boundary conditions change to a no-slip interface (denoted by a solid line), at which the velocity vanishes. 
Using analytical solutions we show that the no-slip interface does not respect bulk-boundary correspondence and supports instead only 2 (two) interface modes. 

In this Letter, we show that this system exhibits \emph{flux trapping} via a mismatch between incoming and outgoing modes. 
We show that, this mechanism implies the existence of non-Hermitian terms for systems described by Hermitian equations, and can be used to design novel behaviour for topological modes. 
An example of such behaviour is \emph{complete absorption}, illustrated in Fig.~\ref{fig:1}(b--d). 
A topologically protected mode propagating along the fluid-fluid interface vanishes once it reaches the change in boundary. 
In a Hermitian system, such energy dissipation is prohibited, and no consistent solution exists. 
To resolve this apparent paradox, we are forced to include non-Hermitian dissipative viscosity in this topological theory.  
We first demonstrate how we achieve flux-trapping by the simultaneous breaking of time-reversal and bulk-boundary correspondence.

\section{II. Topological flux trapping}
\begin{figure}[t]
    \includegraphics[width=\linewidth]{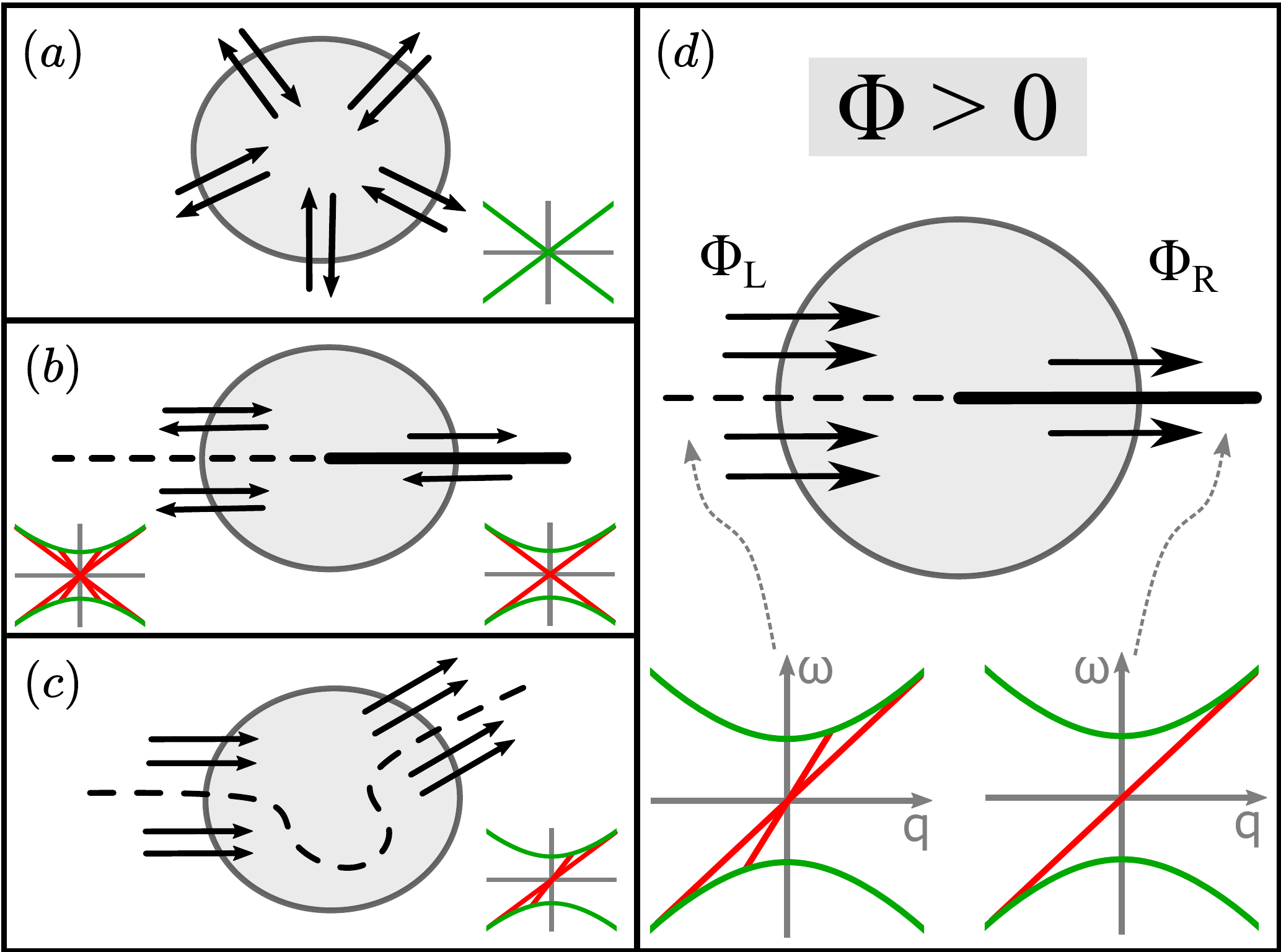}
    \caption{
    Mode absorption and non-Hermiticity. In a region of space (grey circle), we count the number of in- and out-flowing modes (black arrows) and apply the pigeonhole principle. 
    Insets in (a-d) show different bulk (green) and interface (red) dispersion relations.
    (a) Ungapped time-reversal-symmetric system. Every in-flowing mode is paired with a mode flowing out by time-reversal symmetry (TRS). The net mode flux $\Phi$ is zero.
    (b--d) Gapped systems, for which $\Phi = \Phi_L + \Phi_R$, where $\Phi_{L/R}$ counts modes at the two points where the interface (dotted/solid lines) crosses the region's boundary. 
    (b) Gapped system with TRS.
    $\Phi_L=\Phi_R=0$, even if BCs change the number of interface modes.
    (c) System for which TRS is broken,  but bulk-interface correspondence holds. Topological protection guarantees one out-flowing mode for each in-flowing mode, even along complex interfaces, $\Phi_L+\Phi_R =0$.
    (d) When both TRS and bulk-interface correspondence are broken, the in-flowing modes can outnumber the out-flowing modes, so that $\Phi>0$ and non-Hermiticity is guaranteed.
    }
    \label{fig:2}
\end{figure}

Such mode flux trapping [e.g., Fig.~\ref{fig:1}(b--d)] occurs when two exotic conditions are satisfied: the system must \emph{(i)}~break time-reversal symmetry and \emph{(ii)}~violate bulk-interface correspondence.
(We assume that the equations of motion are linear and have wave-like solutions.)
\emph{(i)}~Time-reversal symmetry (in this case equivalent to reciprocity) implies that for every right-moving wave at wavevector $\bm q$ and frequency $\omega(\bm q)$ there exists a left-moving wave with equal frequency $\omega(- \bm q)$. Then, through every boundary there is an equal number of modes flowing in each direction, Fig.~\ref{fig:2}(a--b). 
Therefore, the net number of modes going into any region is zero. \emph{(ii)}~In non-reciprocal systems which satisfy bulk-interface correspondence, the net flux of modes into a region is still zero. This is because topological protection guarantees that for each incoming wave, there exists an outgoing wave along the same boundary at the opposite end, Fig.~\ref{fig:2}(c).

Chiral active fluids satisfy both of these conditions and can have the interface mode structure depicted in Fig.~\ref{fig:2}(d). As a consequence, more modes enter a region than can leave.
The net flux of modes $\Phi$ into a region can be evaluated using the expression
\begin{equation}
    \Phi = \
    \sum\sgn(\bm q\cdot\bmh n),
\end{equation}
which sums the signed number $\sgn(\bm q\cdot\bmh n)$ of modes flowing into a region over all of the boundary elements. In analogy with typical expressions for a flux, the quantity $\sgn(\bm q\cdot\bmh n)$ acts as a vector field dotted into a surface normal. However, this quantity can only be an integer because it corresponds to a mode \emph{count} for flow either into (contributing flux $1$) or out of (contributing flux $-1$) the region.
The mode flux $\Phi$ for the geometry in Fig.~\ref{fig:2}(d) is positive and can be evaluated as the sum of fluxes at the two points $L/R$ where the interface crosses the boundary of the region
\begin{equation}
    \Phi = \Phi_L + \Phi_R = \left(N_L^{in} + N_R^{in}\right) - \left(N_L^{out} + N_R^{out}\right), 
\end{equation}
where $N^{in/out}_{L/R}$ is the number of modes flowing into/out of a region through the point $L/R$.

We now show that for traveling waves, a positive mode flux (i.e., $\Phi > 0$) necessitates non-Hermitian physics. 
We do so by defining an $N^{in} \times N^{out}$ rectangular extension $\tilde{S}$ of the scattering matrix that maps incoming waves into outgoing waves, assuming no bound states. 
For $\Phi > 0$, a theorem of linear algebra states that the rectangular matrix $\tilde{S}$ must map at least $\Phi$ independent modes to zero~\footnote{Modes mapped to zero are called the nullspace of $\tilde{S}$, and the dimensionality of the nullspace, called the nullity, must be at least $\Phi$. Also note the distinction with mapping incoming waves into bound states, in which the scattering matrix remains unitary.}. 
This observation precludes a description of the dynamics in terms of a purely Hermitian Hamiltonian $\cal{H}$, which would instead imply a (typical) unitary scattering matrix $S \equiv e^{i \cal{H}}$ that cannot map any nonzero modes to zero.
This extension of scattering matrices outside the unitary group is typical of non-Hermitian dynamics~\cite{rotter2009non,longhi2010optical,novitsky2020unambiguous,achilleos2017non,feng2017non}, non-reciprocal cavities~\cite{mann2019nonreciprocal} and systems defined by information loss, including black holes~\cite{polchinski2017black}.
In summary, the topological physics associated with a positive mode flux $\Phi$ needs to be regularized by dissipative non-Hermiticity. 

As a toy model for regularization by dissipation, consider a reduced equation for a single chiral edge mode $\psi$ in one spatial dimension $x$, $(\partial_t - \partial_x) \psi = \nu \partial_x^2 \psi$ with a nonzero initial condition for $\psi(x < 0)$ and the boundary condition $\psi(0) = 0$, which ensures that $\psi$ ``vanishes'' for positive $x$. 
Around the origin, $\Phi > 0$, and no bound modes are allowed. For $\nu = 0$, the solution $\psi(x - t)$ cannot be valid for all time. 
By contrast, for $\nu > 0$, the boundary condition remains compatible with the chiral edge mode.

\section{III. Hydrodynamic theory with vanishing modes}

One minimal two-band model for violating bulk-boundary correspondence is a compactified Dirac cone~\cite{Tauber2020}.
We instead demonstrate the flux-trapping mechanism using the hydrodynamic description of chiral active fluids~\cite{souslov2019topological}:
\begin{align}\label{eq:pde}
\begin{split}
    \partial_t \bm v +\nabla \rho &= -s (m+\nabla^2)\bm v^* + \nu \nabla^2 {\bm v},
    \\[\jot]
     \partial_t \rho + \nabla\cdot \bm v &= D \nabla^2 {\rho},
\end{split}
\end{align}
where $\bm x  \equiv (x,y)$, $\bm v \equiv (u,v)$, and $\bm v^* \equiv (-v,u)$, with an interface at $y = 0$ with $s = \sgn(y)$. The linearized Eqs.~(\ref{eq:pde}) describe the time-evolution of fluid density $\rho_0+\rho(\bm x, t)$ and flow velocity $\bm v(\bm x, t)$.
The left-hand side of Eqs.~(\ref{eq:pde}) describes sound waves in a simple fluid.

The term $s (m+\nabla^2)\bm v^*$ in Eqs.~(\ref{eq:pde}) describes the hydrodynamic effects due to active rotations of individual particles and active flows.
Within this chiral active fluid, each particle, with moment of inertia $I$, spins at rate $\Omega_A$ and feels a transverse body force $\bm F = \rho_0 \Omega_B \bm v^*$. 
The fluid stress tensor is $\sigma_{ij} = -p\delta_{ij}+\rho_0\nu^o\,\partial_iv^*_j$, where $\nu^o=I\Omega_A/(2\rho_0)$ is the odd (or Hall) viscosity. 
Odd viscosity is a transverse non-dissipative response~\cite{Avron1998,Kaminski2014} in chiral active fluids~\cite{banerjee2017odd,soni2019odd,markovich2020odd}, quantum fluids and plasmas~\cite{Avron1995,Tokatly2007,Read2011,Offertaler2019,Korving1966}
as well as rotated gases~\cite{Knaap1967}. Without odd viscosity, Eqs.~(\ref{eq:pde}) also describe ocean waves near the equator or next the shore~\cite{Rosenthal1965,Matsuno1966,Yanai1966,delplace2017topological}.

Here, we treat the magnitudes of the intrinsic rotation rate $\Omega_A$ and cyclotron frequency $\Omega_B$ as constants set by active forces. 
For example, assuming our particles carry an electric charge and a magnetic moment, $\Omega_A$ ($\Omega_B$) could be set by a magnetic field rotating in (pointing perpendicular to) the plane~\cite{soni2019odd,vanZuiden2016,souslov2019topological}. 
Further, we assume $\Omega_A$ and $\Omega_B$ have the same sign $s = \sgn(y)=\pm1$. 
The sign change across the interface may be induced  by discontinuously varying the external fields. 
Such interfaces also arise due to segregation of oppositely rotating chiral active particles~\cite{nguyen2014emergent,yeo2015collective,deljunko2018energy,scholz2018rotating}.
In Appendix A, we detail how Eqs.~(\ref{eq:pde}) describe sound waves in a chiral active fluid with one free parameter $m=|I\Omega_A\Omega_B|/(2\rho_0c^2)$ after non-dimensionalizing in the following units:
\begin{table}[h!]
    \centering
    \begin{tabular}{
    @{\hspace{5pt}}c@{\hspace{5pt}}|
    @{\hspace{5pt}}c@{\hspace{5pt}}|
    @{\hspace{5pt}}c@{\hspace{5pt}}}
        %\multicolumn{3}{c}{Units} \\[2pt]
        Density & Length & Time  \\\hline
        $\rho_0$ & $I \Omega_A/(2c\rho_0)$ & $I \Omega_A/(2 c^2 \rho_0)$
    \end{tabular}
\end{table}

In Eqs.~(\ref{eq:pde}), we have included two non-Hermitian terms: a dissipative viscosity $\nu$ and a diffusion coefficient $D$. 
We proceed to show by contradiction that such dissipative terms must be included for the equations to be well posed.
We first assume that $\nu = D = 0$ and provide details in Appendices B-F for solutions to the equations along the interfaces shown in Fig.~\ref{fig:1}.
Bulk solutions to Eqs.~(\ref{eq:pde}) can be decomposed into three bands $\omega(\bm q)$.
These bands are characterized by topological invariant called Chern numbers $C$ because \emph{(i)} the equations are gapped at low frequencies, and no bulk waves exist in the gap~\cite{Shankar2017,souslov2019topological,Tauber2019,Tauber2020};
\emph{(ii)} odd viscosity compactifies the fluid's acoustic spectrum~\cite{Fukui2012,Silveirinha2016,bal2019continuous}.
For Eqs.~(\ref{eq:pde}), $C_s = \sgn(\Omega_A)+\sgn(\Omega_B) = 2s$ (see Appendix A), only defined when both $\Omega_A$ and $\Omega_B$ are nonzero. 
The standard bulk-interface correspondence would suggest the existence of $C_+-C_- = 4$ modes at any interface between systems with opposite values of $s$. 
Intuitively, this bulk-\emph{interface} correspondence suggests the gluing of two bulk-\emph{edge} correspondences glued together.

By characterizing waves at interfaces, we show that this bulk-interface correspondence does not hold for no-slip boundary conditions.
At an interface, we find solutions of the form $\bm\psi\equiv(\rho,u,v)^T = \bm{\Psi}_\ud e^{i(\omega t - qx)} e^{-\kappa_\ud|y|}$, assuming $\omega < m$ (i.e., the frequency is in the gap set by $\Omega_B$) and $m < 1/4$. 
For each halfspace, which we label $\uparrow$ and $\downarrow$, we find two solutions, which we denote by $\bm\psi_\ud^\pm$.
The general solution to Eq.~(\ref{eq:pde}) has the form $\bm\psi = 
A_\uparrow^+\bm\psi_\uparrow^+ + 
A_\uparrow^-\bm\psi_\uparrow^- + 
A_\downarrow^+\bm\psi_\downarrow^+ + 
A_\downarrow^-\bm\psi_\downarrow^-
$, where the amplitudes $A_\ud^\pm$ must be chosen to satisfy the interface boundary conditions (BCs).
The BCs conserve energy if $\bmh{n}\cdot\bm\sigma\cdot\bm v|_{y\downarrow0} = \bmh{n}\cdot\bm\sigma\cdot\bm v|_{y\uparrow0}$, as we review in Appendix A.
Two such interface conditions are (see Figs.~1,3):
\begin{align}\label{eq:BCs}
\begin{split}
    \text{Fluid-Fluid (FF):}\quad  & 
    0 \!=\! [u] \!=\! [v] \!=\! [\sigma_{21}] \!=\! [\sigma_{22}] \\[\jot]
    \text{No-Slip (NS):}\quad  & 
    0 \!=\! [u] \!=\! [v] \!=\! \hspace{7pt}\bar{u} \hspace{7pt} \!=\! \hspace{7pt}\bar{v} 
\end{split}
\end{align}
where we denote by $[u] = u(0^+)-u(0^-)$ the jump across (and by $\bar{u} = (u(0^+)+u(0^-))/2$ the average of) the velocities and stresses at $y=0$.
As usual, at a fluid-fluid interface the stress is continuous and at a no-slip interface, the velocity vanishes. 
For each boundary condition, the coefficients $A_\ud^\pm$ of the general solution are related by a single matrix equation $M\bm A= \bm 0$. We find the dispersion curves for the interface modes using $\det M =0$.

\begin{figure}[t!]
    \centering
    \includegraphics[width=.95\linewidth]{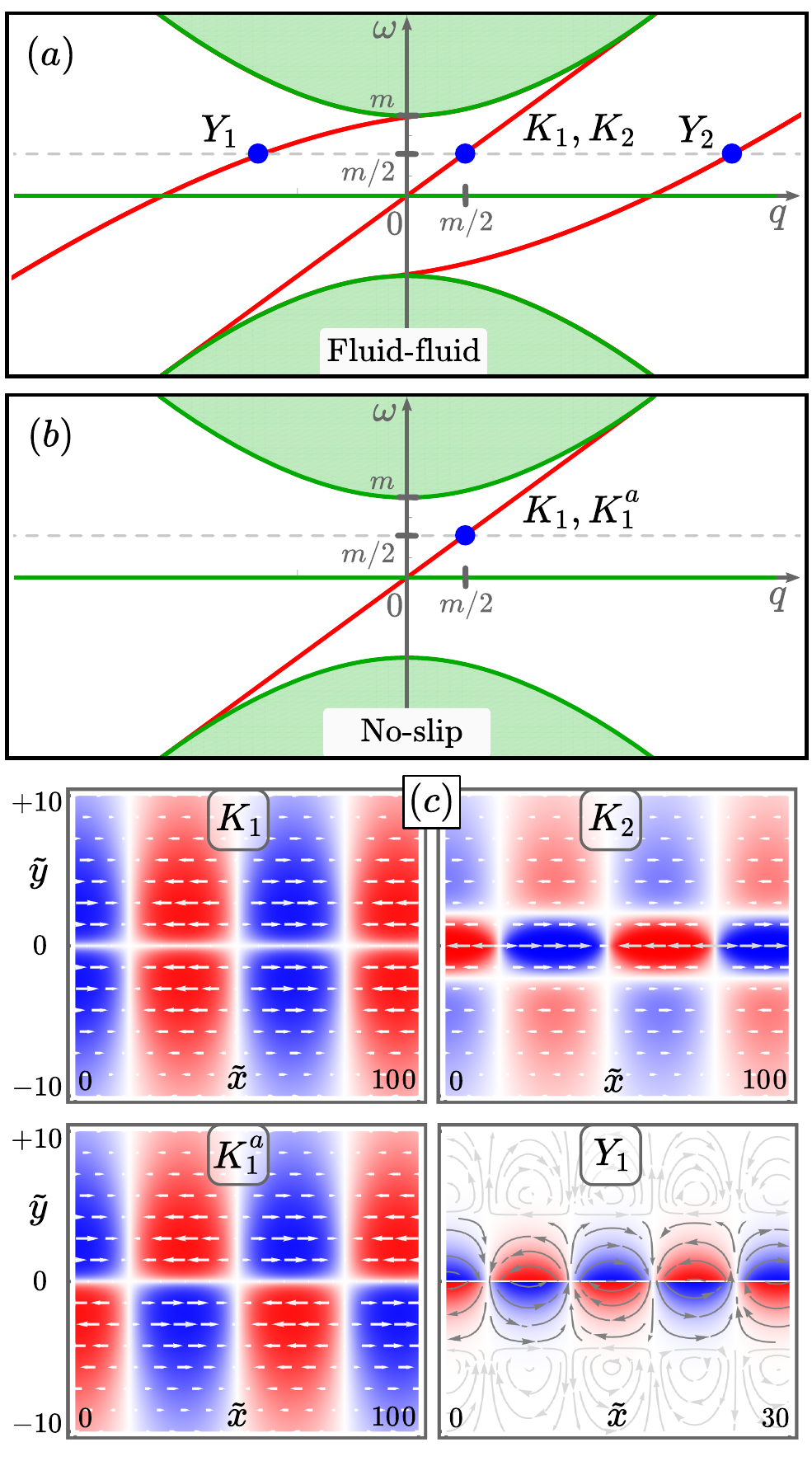}
    \caption{
    Interface solutions for fluids with odd viscosity and no dissipation.
    (a--b) Mode spectrum for two different interfaces between chiral active fluids [Fig.~\ref{fig:1}(a)], showing the dispersion relations for bulk modes in green and for interface modes in red ($m=1/5$). 
    (a) For a fluid-fluid interface, any frequency $\omega$ in the gap corresponds to four interface modes (blue points): two degenerate Kelvin modes ($K_1$,$K_2$) and two non-degenerate Yanai ($Y_1$,$Y_2$) modes. 
    (b) For the no-slip interface, only two degenerate Kelvin modes ($K_1$,$K_1^a$) exist. No Yanai modes are present [c.f.~Fig.~\ref{fig:2}(d)].
    (c) Density (color) and velocity (arrows) of the labeled interface modes.
    The right column ($K_1$ and $K_1^a$) are the only modes on the no-slip interface, while all except the $K_1^a$ mode are present on the fluid-fluid interface.
    The fourth mode on the fluid-fluid interface is the Yanai mode $Y_2$, which is similar in shape to the depicted $Y_1$ mode but with a longer wavelength.
    }
    \label{fig:3}
\end{figure}

Both interface conditions have well posed solutions, even without dissipation.
For the fluid-fluid interface, we find four solutions at each frequency $\omega$ in the gap, in agreement with bulk-interface correspondence (Fig.~\ref{fig:3}). 
These solutions fall into one of two categories: \emph{(i)} two so-called Kelvin modes with linear dispersion $\omega=q$ and horizontal ($v=0$) flow; and \emph{(ii)} two Yanai modes with $|\omega|\neq q$ and purely rotational ($u \propto i v$) flow. 
For Kelvin modes, the explicit solutions (derived in Appendices B-F) are $(\rho,u,v)_{K_{1,2}} = (1,1,0) K_{1,2}(y)e^{i\omega(x-t)}$, where the $y$-profiles $K_{1,2}(y)$ can be orthogonalized as
\begin{align}
\label{eq:kelvin}
\begin{split}
    K_1(y) = \phantom{\Big[}&\sinh\left(\tfrac{1}{2}\Gamma|y|\right)e^{-\frac{1}{2}|y|},
    \\
    K_2(y) = \Big[&\cosh\left(\tfrac{1}{2}\Gamma|y|\right)-\frac{1}{\Gamma}\sinh\left(\tfrac{1}{2}\Gamma|y|\right)\Big]e^{-\frac{1}{2}|y|}, 
\end{split}
\end{align}
with $\Gamma(\omega) = \sqrt{1-4m+4\omega^2}$. 

By contrast, for the no-slip interface, no Yanai modes exist in the gap and there are again two orthogonal Kelvin modes $\{K_1,K^a_1\}$, where $K^a_1(y) = \sgn(y) K_1(y)$ is the antisymmetrization of $K_1$ about the interface. 
The NS case can be interpreted as two independent copies of a hard edge~\cite{Tauber2020}, which leads to an equivalent basis of solutions $K_1\pm K^a_1$ in either half-space.
This total count of 2 interface modes differs from the jump in Chern number across the interface, which is $4$.
This discrepancy suggests an apparent violation of the usual bulk-interface correspondence.

The extended scattering matrix $\tilde{S}$ for the fluid-fluid to no-slip transition sends $K_1$ to itself and annihilates $K_2$. 
We characterize this scattering via symmetry arguments and finite-element simulations of Eqs.~(\ref{eq:pde}).
For the symmetry argument, consider the parity under reflection $y \rightarrow - y$ for components of each mode:
\begin{table}[h!]
    \centering
    \begin{tabular}{c||c|c||c|c}
        & \multicolumn{2}{c||}{Fluid-fluid} & \multicolumn{2}{c}{No-slip} \\[2pt]
        & $K_1$, $K_2$ & $Y_1$, $Y_2$ & $K_1$ & $K_1^a$ \\\hline
        $\rho,u$ 
        & even & odd & even & odd \\
        $v$ 
        & \hspace{10pt}zero\hspace{10pt} & \hspace{10pt}even\hspace{10pt} & \hspace{10pt}zero\hspace{10pt} & \hspace{10pt}zero\hspace{10pt}
    \end{tabular}
\end{table}

\begin{figure}
    \centering
    \includegraphics[width=.9\linewidth]{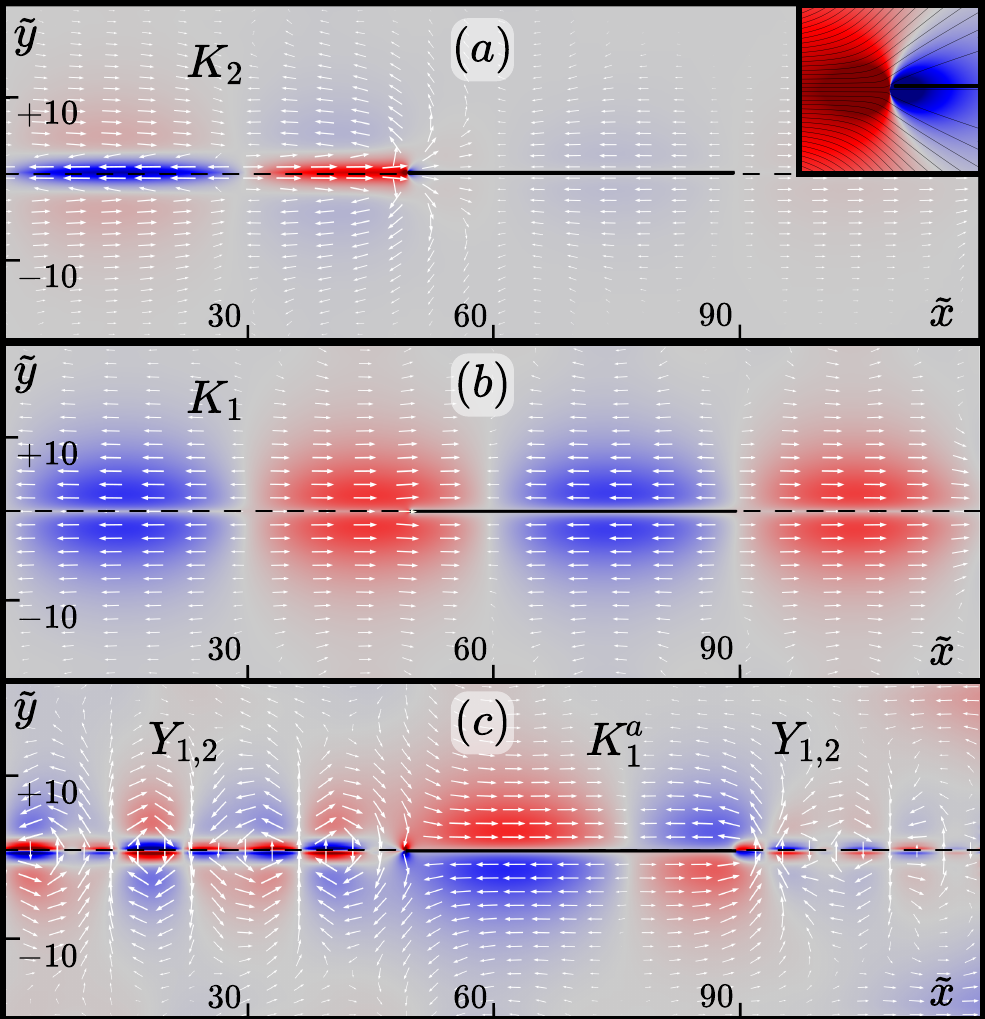}
    \caption{
    Simulations for a change of boundary conditions for fluids with odd viscosity and small dissipation. 
    Density (color) and velocity (arrows) of interface modes coming from the left along the fluid-fluid boundary (dashed line) and interacting with a finite-length segment of no-slip wall (solid black line). 
    (b) The $K_2$ Kelvin wave is orthogonal to all waves supported on the no-slip wall and thus cannot be transmitted. The inset shows a zoom-in, including streamlines, near the change in boundary.
    (c) The $K_1$ Kelvin wave is supported on both interfaces and traverses the no-slip wall unaltered.
    (d) The antisymmetric Yanai waves map to the antisymmetric Kelvin mode $K_1^a$ on the no-slip wall.
    Simulations were performed with parameters $m=0.2$, $D=0$, $\nu/\nu^o = 0.01$.
    }
    \label{fig:4}
\end{figure}

\noindent Because inner products between waves with opposite parity vanish, we decompose $\tilde{S}$ into odd and even parts, i.e., 
$\tilde{S}:\langle Y_1,Y_2 \rangle\to\langle K^a_1 \rangle$
and
$\tilde{S}:\langle K_1,K_2 \rangle\to\langle K_1 \rangle$.
The even-parity sector indeed maps mode $K_1$ onto itself and annihilates the orthogonal mode $K_2$, which we observe in simulations, Fig.~\ref{fig:4}. 
This completes our contradiction: when $\nu = D = 0$, the Hermitian Eqs.~(\ref{eq:pde}) do not allow for mode flux $\Phi$ to be trapped in a region surrounding the change in boundary conditions from a fluid-fluid to a no-slip interface. 
 Applying the results from Ref.~\cite{Tauber2020} shows that a transition from a perfect-slip to a no-slip edge would also lead to topological flux trapping and absorbed modes.
A minimal physical description of the fluid must account for the complete absorption of the mode $K_2$ through non-Hermitian physics.

\section{IV. Energy dissipation in finite time}
\begin{figure}
    \centering
    \includegraphics[width=.95\linewidth]{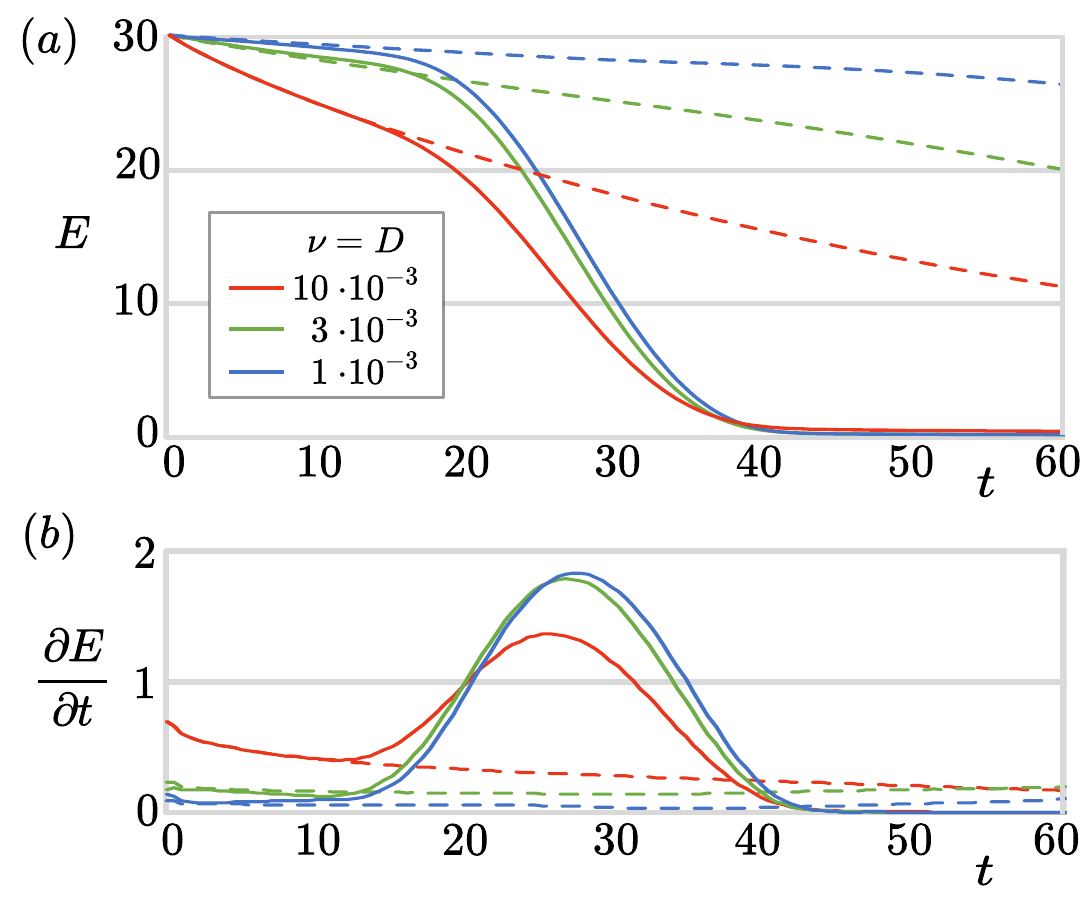}
    \caption{
    Plot of (a) the total energy $E$ and (b) dissipation rate $\partial E/\partial t$ of a single pulse composed of the $K_2$ Kelvin wave, for various values of $\nu$ ($ = D$). The pulse travels at constant speed along the fluid-fluid interface with (solid lines) or without (dashed lines) a change in boundary conditions to a no-slip wall. In the former case, the front (tail) of the wave hits the no-slip wall at t=10 (t=40).
    }
    \label{fig:5}
\end{figure}

To regularize our dynamics we add non-zero dissipative terms back into Eq.(\ref{eq:pde}). 
These terms include the viscosity $\nu$, which enters the velocity equation, and the diffusion coefficient $D$, which enters the continuity equation.
Without a change in boundary conditions, these dissipative effects lead to a decrease in energy through the mechanism of attenuation. 
With a change in boundary conditions, these waves can in addition lose energy by being absorbed.
From our results, we conclude that this absorption happens much faster than the attenuation, as shown in Fig.~\ref{fig:5}. Strikingly, absorption dissipates all of the energy by the time the tail of the wave packet reaches the change in boundaries, irrespective of the values of dissipative parameters $\nu$ and $D$. 
We deduce that the absorption time of a trapped mode scales only with the wave period, giving an absorption timescale of $\tau_a\sim\omega^{-1}$ (i.e., the wavelength $1/q$ over phase velocity $\omega/q$). 
By contrast, the attenuation timescale $\tau_S$ for any transmitted mode was calculated by Stokes to be $\tau_S \sim c^2/(\omega^2 \nu)$~\cite{LandauVI}.
Because $\tau_S$ diverges when dissipation vanishes, for small viscosities absorption occurs much faster than attenuation, i.e., $\tau_a \ll \tau_S$.

Further, we stress that the dissipation rate due to absorption remains finite in the limit of vanishing dissipative parameters $\nu$ and $D$, as seen in Fig.~\ref{fig:5}. This introduces a length scale for the region in which dissipation takes place. The dynamical equation for the energy density can be derived from Eq.~(\ref{eq:pde}) to be
\begin{equation}\label{eq:energydensity}
    \partial_t\left(\rho^2+\Vert\bm v\Vert^2\right) = 
        -\nu\left( \Vert\bm\nabla u\Vert^2  
        +\Vert\bm\nabla v\Vert^2  \right)
        -D\Vert\bm\nabla \rho\Vert^2,
\end{equation}
where we have left out the boundary term. Hence, for the left-hand side of Eq.~(\ref{eq:energydensity}) to remain finite as $\nu$ and $D$ tend to zero, the gradients of $\bm v$ and $\rho$ have to scale with $\nu^{-1/2}$ and $D^{-1/2}$ respectively. 
The lengthscale $\ell \sim \sqrt{\nu/\omega}$ is then set by  balancing the incoming energy flux $O(\omega |{\bm v}|^2)$ against dissipation $O(\nu |{\bm v}|^2/\ell^{2})$ due to velocity gradients ${\bm v}/\ell$.
Even for vanishingly small but nonzero dissipative terms, the mode $K_2$ remains trapped inside this region, leading to complete absorption. 

\section{V. Outlook and Conclusions}
\begin{figure}
    \includegraphics[width=.7\linewidth]{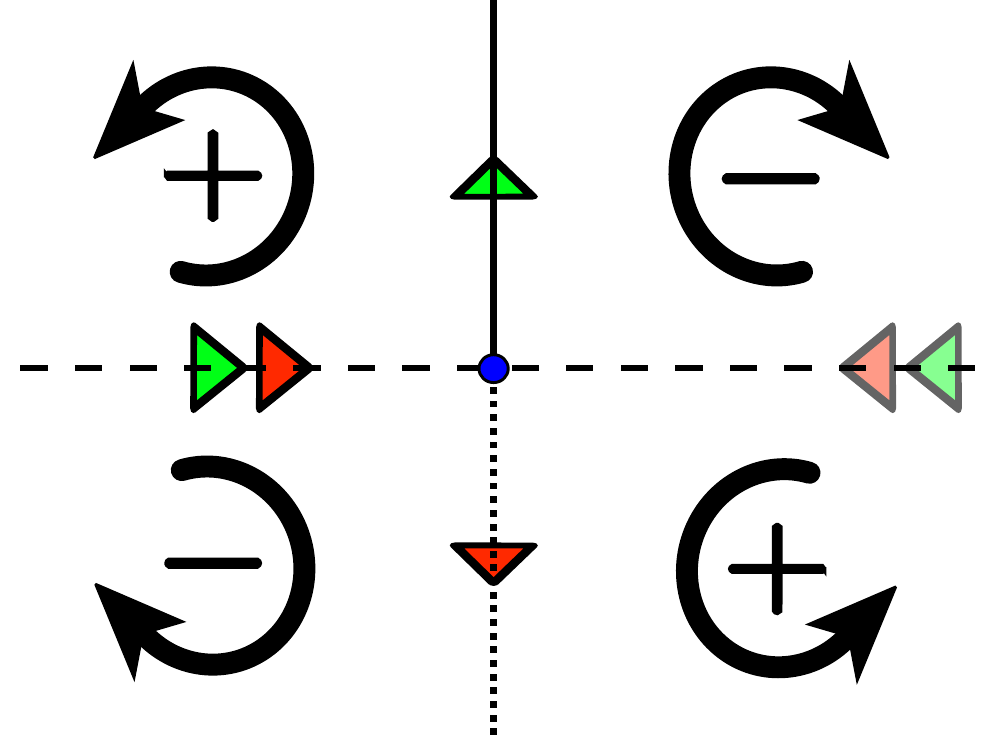}
    \caption{
    Schematic of a beam-splitting setup with topological interfaces. 
    The infinite plane is divided into four quadrants with alternating signs for the Chern numbers originating, e.g., from microscopic rotations as illustrated by the arrows.
    The horizontal interface supports two independent unidirectional modes (indicated by the green and red triangles), which can flow either from the right or from the left in accordance with the bulk-boundary correspondence principle. 
    Each of the two different vertical interfaces break bulk-boundary correspondence by supporting only one of the two incoming unidirectional modes.
    }\label{fig:6}
\end{figure}

One potential application for these interface modes is in designing beam splitters. 
While applications of periodic lattices for splitting topological interface modes have been considered~\cite{makwana2018designing}, we consider continuum systems instead.
The non-zero mode flux particular to continuum systems suggests
a distinct design for topological beam splitters.
We envision the specific setup in Fig.~\ref{fig:6}, with three interfaces between topological fluids meeting at a single junction. 
The horizontal interface satisfies bulk-boundary correspondence and supports two independent modes, both of which propagate towards the junction.
Each of the two different vertical interfaces with broken bulk-boundary correspondence support only one mode, which propagates away from the scattering point.
A challenge is to design interface conditions for which the two outgoing waves form an orthogonal basis for the incoming waves. Then, any incoming signal will be perfectly split between the two outgoing channels.
The advantage is that one can tune the relative power in the outgoing channels by tuning the shape of the incoming signal instead of tuning the structure.

\begin{figure}
    \includegraphics[width=.9\linewidth,height=.43\linewidth]{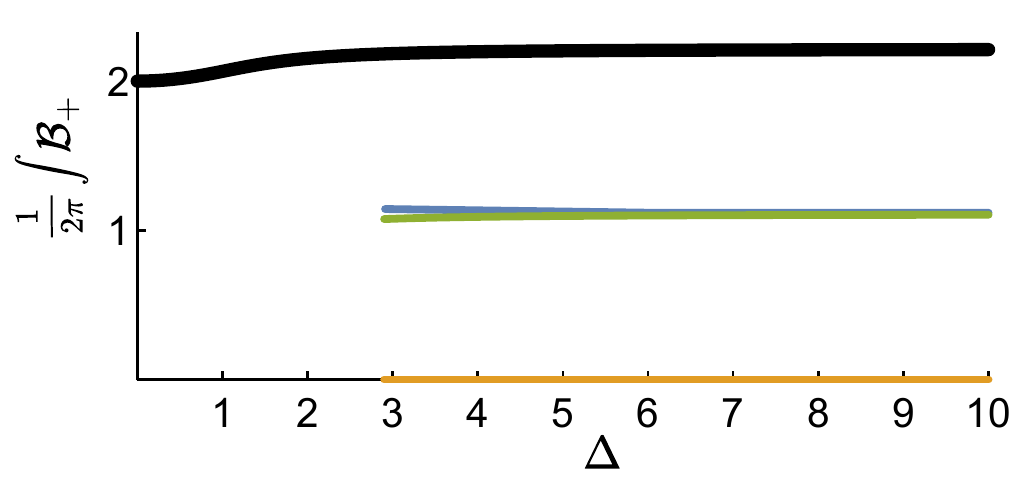}
    \caption{
    The total Berry curvature (black) integrated over the two-dimensional space of wavenumbers $\bm q$, plotted against the dissipative difference parameter $\Delta = D-\nu$. Around $\Delta = 2.8$ we can split the integral in three sections as seen from the origin of the $\bm q$-plane: before (blue), in between (yellow) and beyond (green) the exceptional points. The blue and green lines are directly on top of each other, indicating that total Berry curvature is equally shared between the two sections outside the exceptional points. 
    }\label{fig:7}
\end{figure}

In addition to practical applications, this systems exhibits a surprising interplay between Hermitian
and non-Hermitian physics.
At larger dissipation, a pair of exceptional points emerges for which the non-Hermitian topology~\cite{brandenbourger2019non,ghatak2020observation,Rosa2020,helbig2020generalized,scheibner2020odd,schomerus2020nonreciprocal,Budich2019,gong2018topological,yoshida2019exceptional,shen_topological_2018,Hatano1996,Bender1998,lee2019topological} could interplay with the Hermitian topological features described in this manuscript. 
At exceptional points, bands cross and we expect that Chern numbers would not be well defined as the integral of the Berry curvature over the top band. 
In Appendix G, we detail how in our system we can still uniquely assign Berry curvature to each band.
We observe that, even though the Berry curvature contains singularities, its integral converges and remains close to the Hermitian value, see Fig.~\ref{fig:7}.
We note that this total curvature is equally shared between the regions separated by the exceptional points.
In this way, the total Berry curvature may still carry topological information and coexist with the winding number carried by the exceptional points. This provides a platform to explore topological states that simultaneously exhibit both Hermitian and non-Hermitian topological invariants.

In conclusion, we have demonstrated that when both time-reversal symmetry and bulk-interface correspondence are broken, topological modes can flow into regions from which they cannot flow out. 
This shows a minimal description for (topologically Hermitian) chiral edge states must include (non-Hermitian) dissipative mechanics.
To illustrate this phenomenon, we considered an active-fluid system that contains a region with  positive mode flux. 
In this region, we find that some modes get completely absorbed, even when dissipative terms are vanishingly small. We construct mathematical expressions for these absorbed modes.
Previously, chiral edge modes have been developed to construct robust waveguides for the conduction of electric current, light, or sound~\cite{von_klitzing_quantized_1986,thouless_quantized_1982,haldane_model_1988,hasan2010colloquium,rechtsman_photonic_2013,lu_topological_2014,ozawa2019topological,Nash2015,wang_topological_2015,Wang2015a,Kariyado2015,Shankar2017,Dasbiswas2017,mitchell_amorphous_2018,souslov2019topological,susstrunk_classification_2016,bertoldi2017flexible,Shankar2020}. 
Building on our hydrodynamic model, topology could also be exploited to design perfect absorbers to completely shield against incoming signals. 
Allowing the no-slip interface to vibrate or move could open further avenues for harvesting energy from and propulsion by these absorbed waves.

%------------------------------------------------------------------------------------------
\acknowledgements{\section{Acknowledgments} P.A.M.~gratefully acknowledges support through the Royal Society project IEC\textbackslash{}R2\textbackslash{}170195. A.S.~gratefully acknowledges the support of the Engineering and Physical Sciences Research Council (EPSRC) through New Investigator Award No.~EP/T000961/1.
}

\appendix
\subsection{Appendix A: Odd viscosity in fluids of spinning particles}
In this section, we review the hydrodynamic origins of Eqs.~(\ref{eq:pde}) as a description of acoustic waves in two-dimensional chiral active fluids~\cite{banerjee2017odd,souslov2019topological,markovich2020odd}. 
Consider an ideal fluid of spinning particles, subject to a Coriolis (equivalently, Lorentz) body force. The equation of motion in terms of the mass density $\rho(x,y,t)$ and fluid flow $\bm v(x,y,t)$ is
\begin{equation}\label{eq:fluideqv}
    \partial_t(\rho \bm v) +\nabla\cdot\left[\rho\bm v\otimes\bm v - \bm{\underline\sigma}\right] 
    = \Omega_B\rho\bm v^*,
\end{equation}
where the $*$ operator acts on vectors $\bm v=(u,v)$ as $(u,v)^*=(-v,u)$, and the stress tensor is given by
\begin{equation}\label{eq:stress}
    \sigma_{ij} = -p\delta_{ij} + \eta^o (
    \partial_i v^*_j + 
    \partial^*_i v_j)  
    + \eta (
    \partial_i v_j + \partial_j v_i - \partial_k v_k \delta_{ij} ),
\end{equation}
where $p$ is the pressure, $\eta$ is the (dynamic) dissipative shear viscosity, and $\eta^o$ is the (dynamic) odd viscosity, with kinematic viscosities used in the text defined by $\nu^o = \eta^o/\rho_0$ and $\nu = \eta/\rho_0$. 
Although we do not include the dissipative \emph{bulk} viscosity term $\zeta \partial_k v_k \delta_{ij}$ in the stress tensor, qualitatively this term would not affect our conclusions. 
Both the dissipative and odd viscosities arise from the same viscosity tensor $\eta_{ijkl}$, defined as rank-4 tensor that linearly relates the rank-2 stress tensor to the rank-2 strain rate tensor, $\sigma_{ij} = \eta_{ijkl} \partial_k v_l$. 
The difference is that the part of the viscosity tensor $\eta^e_{ijkl}$ containing dissipative viscosity is \emph{even} under time-reversal symmetry and therefore (by Onsager reciprocity relations) has to satisfy $\eta^e_{ijkl} = \eta^e_{klij}$, i.e., the dissipative viscosity enters the even part of the viscosity tensor under the exchange $ij \leftrightarrow kl$. 
By contrast, the part of the viscosity tensor $\eta^o_{ijkl}$ that contains odd viscosity is \emph{odd} under time-reversal symmetry and therefore has to satisfy $\eta^o_{ijkl} = - \eta^o_{klij}$, i.e., the odd viscosity enters the odd part of the viscosity tensor under the exchange $ij \leftrightarrow kl$. 
A peculiar property that arises from the symmetry of odd viscosity is that this term does not lead to energy dissipation, as can be checked directly by evaluating the entropy production rate as $\tfrac{1}{2} \eta^o_{ijkl} \partial_i v_j \partial_k v_l = 0$. 
Because odd viscosity must break time-reversal symmetry, this term naturally arises in systems with an external magnetic field or rotation. 
In chiral active fluids, the intrinsic rotation of each individual particle gives rise to this term.

General arguments for the value of odd viscosity can be made in cases in which the spinning particles composing the fluid are weakly interacting, $\eta^o = \pm \ell/2$, where $\ell$ is the density of the intrinsic angular momentum  for the individual particles. 
The $\pm$ sign in the relation depends on the microscopic origins of the effect~\cite{Wiegmann2014,banerjee2017odd,markovich2020odd}. 
For some  chiral active fluids with inertia, $\ell = I \Omega_A$, where each particle rotates with intrinsic angular velocity $\Omega_A$ with $I$ the individual particles' moment of inertia times the fluid density~\cite{banerjee2017odd, souslov2019topological}. 
In a variety of fluids, odd viscosity arises due to interparticle interactions~\cite{Han2020} and dissipative processes~\cite{souslov2020anisotropic}, rendering it a distinct transport coefficient independent of the intrinsic angular momentum $\ell$. 
For simplicity, we choose to define the chiral terms via the frequencies $\Omega_A$ and $\Omega_B$, which act as proxies for the magnitude of the chiral body force and odd viscosity, respectively. 
We will consider only cases where in the bulk of the fluid these two quantities have the same non-zero sign, i.e., $\Omega_A = s|\Omega_A|$ and $\Omega_B = s|\Omega_B|$ with $s=\pm1$. 

We linearize Eq.~(\ref{eq:fluideqv}) together with the continuity equation $\partial_t\rho +\nabla\cdot(\rho\bm v)=0$ around $\rho=\rho_0$ and $\bm v=\bm 0$ to obtain Eq.~(\ref{eq:pde}) in the main text. 
We use $\eta^o = I \Omega_A/2$, $c^2= \partial p/\partial\rho$, and choose units of mass density in terms of $\rho_0$, length in terms of  $I \Omega_A/(2c\rho_0)$, and time in terms of $I \Omega_A/(2 c^2 \rho_0)$. 
In these units, (kinematic) dissipative viscosity $\nu$ is measured in units of the (kinematic) odd viscosity $\nu^o$. 
This non-dimensionalized system is controlled by the discrete constant $s=\pm1$ and the positive constant $m = |I\Omega_A\Omega_B|/(2\rho_0c^2)$, which is assumed to satisfy $m<1/4$ to ease later computation. 

We note that the linearized equations conserve the (rescaled) energy $E = \rho^2 + \Vert\bm v\Vert^2$, which can be seen by deriving the energy continuity equation
\begin{equation}\label{eq:Energyconservation}
    \partial_t\int_V E\, dA = 
    -2\int_{\partial V}\bmh n\cdot\bm{\underline\sigma}\cdot\bm v\,d\ell.
\end{equation}
Suppose that we have two regions $V_1$ and $V_2$ which share a piece of their boundary $\partial V_{12}$ ($=\partial V_1\cap\partial V_2$). 
For the energy $E$ to be conserved, we need the flux \emph{out} of $V_1$ to equal the flux \emph{into} $V_2$ and vice-versa. 
That is, the flux $\bmh n\cdot\bm{\underline \sigma}\cdot\bm v$ needs to be continuous at $\partial V_{12}$. 
There are two natural ways to guarantee that to $\bmh n \cdot \bm{\underline\sigma} \cdot \bm v$ is continuous, namely (1) Stress-continuity: continuous $\bm v$ and continuous $\bmh n\cdot\bm{\underline\sigma}$; or (2) Velocity-vanishing: continuous $\bm v$ and $\bm v|_{\partial V_{12}} = 0$.
Case (1) governs the dynamics of the fluid bulk and \emph{fluid-fluid interfaces}. Case (2) governs the dynamics where the fluid touches a \emph{no-slip wall}.

\subsection{Appendix B: Bulk solutions}
In this section, we review the evaluation of Chern number within the dispersion relation associated with Eqs.~(\ref{eq:pde})~\cite{souslov2019topological,Tauber2019,bal2019continuous}.
We can write Eqs.~(\ref{eq:pde}) as a Hamiltonian equation $-i\partial_t\bm\psi = H_s(i\partial_x,i\partial_y)\bm\psi$ in terms of $\bm\psi=(\rho, u, v)^T$, where
\begin{equation}
    H_s(i\bm\nabla) = 
    \begin{pmatrix}
        0 & i\partial_x & i\partial_y \\
        i\partial_x & 0 & -i s\left(m-(i\bm\nabla)^2\right) \\
        i\partial_y & + i s\left(m-(i\bm\nabla)^2\right) & 0
    \end{pmatrix}
\end{equation}
In the bulk, the Hamiltonian commutes with all time and space derivatives, so we may assume bulk waves of the form $\bm\psi=\bm\Psi e^{i(\omega t-\bm q\cdot\bm x)}$. 
The Hamiltonian equation then reduces to a family of eigenvalue equations $H_s(\bm q)\bm\Psi = \omega\bm\Psi$, with eigenvalues:
\begin{equation}\label{eq:bulkdispersion}
    \omega_0(\bm q) = 0 \quad\text{or}\quad \omega_\pm(\bm q) = \pm\sqrt{\lambda^2+ q_x^2+q_y^2},
\end{equation}
where $\lambda(\bm q)=m-q_x^2-q_y^2$, and eigenvectors
\begin{equation}\label{eq:bulkeigenvec}
    \bm\Psi_0(\bm q) \propto
    \begin{pmatrix}
        \lambda \\ iq_y \\-iq_x
    \end{pmatrix}
    ,\quad
    \bm\Psi_\pm(\bm q) \propto
    \begin{pmatrix}
        iq_y\lambda + q_x\omega_\pm \\ q_x^2 + \lambda^2 \\ q_xq_y +i\lambda\omega_\pm
    \end{pmatrix}.
\end{equation}
Significantly, this orthogonal basis is well-defined over the compactified plane of wavevectors $\bm q$, for which all of the points at infinity are identified.

The Chern number for the lower band, which determines the number and chirality of modes in systems satisfying bulk-boundary correspondence, can be evaluated from the Berry curvature. 
The Berry curvature of each band in Eq.~(\ref{eq:bulkeigenvec}) is a function of the wavevector $\bm q$ given by
\begin{equation}\label{eq:BerryCurv}
{
{\cal B}_{0,\pm}(\bm q) = i \nabla_{\bm q} \times [\bm\Psi_{0,\pm}^\dagger(\bm q) \cdot \nabla_{\bm q} \bm\Psi_{0,\pm}(\bm q)],
}
\end{equation}
where the dot product contracts eigenvector components, whereas the gradient and curl are vectors in $\bm q$-space.
Here ${\cal B}_0 = 0$ is trivial, leaving ${\cal B}_+ = -{\cal B}_-$. 
Integrating the Berry curvature over all of wavevector space gives the Chern number for the lowest band~\cite{souslov2019topological,bal2019continuous}, 
\begin{equation}
\mathcal{C}_- = \frac{1}{2\pi}\int {\cal B}_- \,\,d{\bm q} 
    = \sgn(\Omega_A) + \sgn(\Omega_B)
    = 2s,
\end{equation}
which determines the number of edge states when the bulk-boundary correspondence holds.

\subsection*{Appendix C: Halfspaces}
Now suppose our space is partitioned by a boundary along $y=0$ into an upper ($y>0$) and lower ($y<0$) halfspace where, respectively, $s_\uparrow=+1$ and $s_\downarrow=-1$. 
Because of the discontinuity in $s$, Eq.~(\ref{eq:pde}) is ill-defined at the interface $y=0$. 
We proceed by solving the PDE on the upper (lower) halfspace separately to obtain general solutions $\bm\psi_\uparrow$ ($\bm\psi_\downarrow$). 
We will stitch these solutions together with interface conditions along $y=0$.

Since the PDE is linear with constant coefficients on each halfspace, we may decompose the solutions in exponentials:
\begin{equation}\label{eq:halfspaceAnsatz}
    \bm\psi_\ud = \bm\Psi_\ud e^{-\kappa_\ud|y|}e^{i(\omega t-q x)}.
\end{equation}
To ensure localization, we require $\mathrm{Re}\,\kappa_\ud>0$. 
We proceed by fixing a pair $(\omega,q)\in\mathbb R^2$, with $\omega$ in the gap, and checking for complex solutions $\bm\Psi_\ud$ and $\kappa_\ud$ at each pair $(\omega,q)$. 
The Hamiltonian equation now reads $H(q,-i s_\ud\kappa_\ud)\bm \Psi_\ud = \omega\bm\Psi_\ud$, with
\begin{equation}\label{eq:boundaryHamiltonian}
    H(q,-is\kappa) = 
    \begin{pmatrix}
        0 & q & -is\kappa \\
        q & 0 & -i s\lambda \\
        -is\kappa & + i s\lambda & 0
    \end{pmatrix},
\end{equation}
and $\lambda = m-q^2+\kappa^2$. 
The characteristic equation $\det[H(q,-i s_\ud\kappa_\ud)-\omega]=0$ is independent of $s_\ud$, thus $\kappa_\ud$ has the same values on the top and bottom halfspaces and we may drop the arrow subscript. 
Assuming $\omega\neq0$, the characteristic equation has four solutions for $\kappa$:
\begin{equation}\label{eq:kappa}
    4\left(\kappa_\pm\right)^2 = 1 \pm 2\Gamma(\omega)+\Gamma^2(q),
\end{equation}
where we use the shorthand $\Gamma(x) \equiv \sqrt{1-4m+4x^2}$, which is real and positive by our assumption $m<1/4$. 
Since solutions to Eq.~(\ref{eq:kappa}) come in $\pm$ pairs, we always find exactly two solutions with $\mathrm{Re}\,\kappa_\pm<0$ that diverge as $y\to\infty$, and two with $\mathrm{Re}\,\kappa_\pm>0$ that are localized. 
In particular, one can show that the right-hand side of Eq.~(\ref{eq:kappa}) is positive for every $\omega$ in the gap. 
Thus, the admissible $\kappa$ (and corresponding $\lambda$) are
\begin{align}
    \kappa_\pm(\omega,q) &= \frac{1}{2}\sqrt{1 \pm 2\Gamma(\omega)+\Gamma^2(q)},\\
    \lambda_\pm(\omega) &= \frac{1}{2}\sqrt{1 \pm 2\Gamma(\omega)+\Gamma^2(\omega)}.
\end{align}
The corresponding eigenvectors on the two half-spaces can then be denoted as
\begin{equation}\label{eq:eigenvec}
    \bm\Psi_\ud^\pm (\omega,q)
    \propto
    \begin{pmatrix}
    \kappa_\pm\lambda_\pm+q\omega \\ 
    \kappa_\pm^2+\omega^2 \\
    i s_\ud\left(
    \omega\lambda_\pm - q\kappa_\pm
    \right)
    \end{pmatrix}
    =
    \begin{pmatrix}
    R^\pm\\
    U^\pm\\
    is_\ud V^\pm
    \end{pmatrix}.
\end{equation}
These can be quickly checked against Eq.~(\ref{eq:boundaryHamiltonian}) using $\kappa_\pm^2 - \lambda_\pm^2 = q^2-\omega^2$. 
The general solutions on the two half-spaces can then be written as
\begin{equation}\label{eq:generalsolns}
    \bm\psi_\ud = \sum_{\pm} A_\ud^\pm \bm\Psi_\ud^\pm e^{-\kappa_\pm|y|}e^{i(\omega t-qx)},
\end{equation}
with four complex coefficients $A_\uparrow^+$, $A_\uparrow^-$, $A_\downarrow^+$ and $A_\downarrow^-$, which have to be fixed by the interface condition.

\subsection*{Appendix D: No-slip interface modes}
The velocity-vanishing condition on the no-slip wall reads $[u]=[v]=\bar u=\bar v=0$. 
Recall that $\bm\psi=(\rho,u,v)$, and that the jump and average of a function $f(y)$ at $y=0$ are defined as $[f]=f_\uparrow(0) + f_\downarrow(0)$ and $2\bar f=f_\uparrow(0) + f_\downarrow(0)$ respectively. 
Substituting Eq.~(\ref{eq:generalsolns}), we can write this condition as a matrix equation
\begin{equation}
    \begin{pmatrix}
    \bm 0 \\ \bm 0
    \end{pmatrix}
    =
    \begin{pmatrix}
     M & - M \\  M &  M
    \end{pmatrix}
    \begin{pmatrix}
    \bm{A}_\uparrow \\ \bm{A}_\downarrow
    \end{pmatrix}
    =
    \begin{pmatrix}
    M(\bm{A}_\uparrow - \bm{A}_\downarrow) \\
    M(\bm{A}_\uparrow + \bm{A}_\downarrow)
    \end{pmatrix},
\end{equation}
on the block vectors $\bm{A}_\ud =(A_\ud^+, A_\ud^-)^T$, where the submatrix $M$ depends on $(\omega,q)$ through $\bm\Psi_\ud^\pm$ as
\begin{equation}
    M = 
    \begin{pmatrix}
        U^+ & U^- \\
        i V^+ & i V^- 
    \end{pmatrix}.
\end{equation}
Thus the interface condition is satisfied if and only if $ M\bm A_\uparrow = M\bm A_\downarrow = \bm 0$. 
In particular, we only find non-zero solutions at those $(\omega,q)$ for which $\det M(\omega,q)=0$.

\begin{figure}[t!]
    \centering
    \includegraphics[width=.45\linewidth]{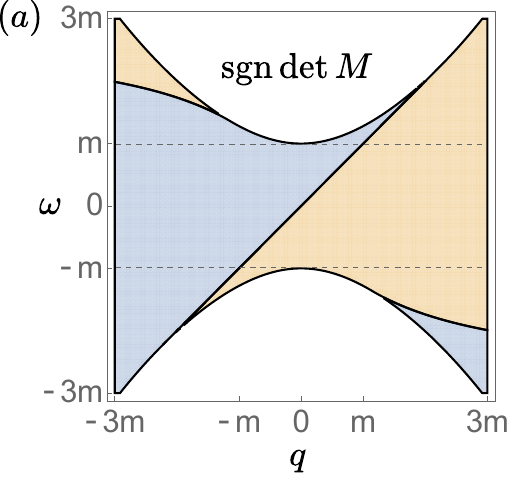}
    \includegraphics[width=.45\linewidth]{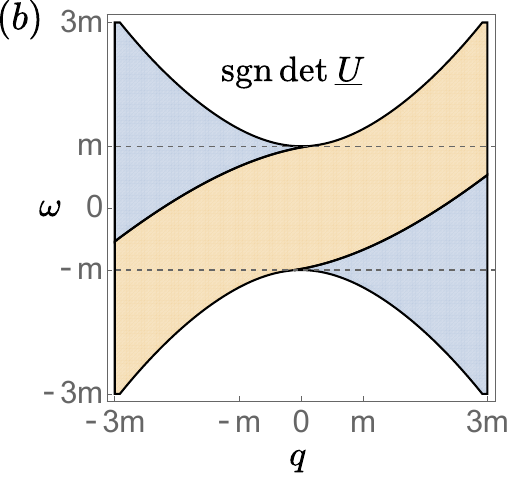}
    \caption{Invertibility of submatrices that determine the interface conditions. 
    The determinants are real between the bulk curves $\omega=\omega_\pm(q)$. 
    Interface waves exist along the curves where the determinant changes from negative (blue) to positive (yellow). 
    (a) For the no-slip interface, $\det M=0$ has only one doubly degenerate Kelvin solution for every $\omega$ in the gap (dashed lines). 
    (b) For the fluid-fluid interface, $\det\underline U=0$ gives two singly degenerate Yanai curves, in addition to the doubly degenerate Kelvin curve $\omega=q$ (not depicted) corresponding to $\det\underline V=0$. 
    }\label{fig:sgndet}
\end{figure}

First let us assume $\omega=q$. Then $\lambda_\pm=\kappa_\pm$ and, by Eq.~(\ref{eq:eigenvec}), $R^\pm = U^\pm\neq0$ and $V^\pm=0$. 
In other words, the velocity normal to the boundary vanishes ($v=0$), and the density equals the tangent velocity ($\rho=u$). 
Since only the second row of $M$ vanishes, the two zero-eigenvectors $\bm A_\ud$ have to be colinear, but we can choose their sign and magnitude independently. 
In particular we may take an orthogonal basis with one even ($\bm A_\uparrow =\bm A_\downarrow$) and one odd solution ($\bm A_\uparrow = -\bm A_\downarrow$) about $y=0$. 
We can write these solutions as $\bm\psi = (1,1,0)^TK(y)e^{-\frac{1}{2}|y|}e^{i\omega( t-x)}$ where the two $y$-profiles are differentiated by
\begin{align}
    K_1 = \sinh\left(\tfrac{1}{2}\Gamma|y|\right)
    \quad\text{and}\quad 
    K_1^a = \sinh\left(\tfrac{1}{2}\Gamma y \right).
\end{align}
These waves with flow parallel to the interface are known as \emph{Kelvin waves}. 
No zero-eigenvectors for $M(\omega,q)$ exist at other $(\omega,q)$ in the gap, as shown in Fig.~\ref{fig:sgndet}(a).

\subsection*{Appendix E: No-slip edge modes}{
Alternatively, we can find the space of modes along the no-slip wall by looking at the modes on either side. 
Since the halfspaces on either side of the wall are decoupled, the mode-space of the hard wall will be the direct sum of the mode-spaces on either side. 

Looking at the top halfspace, we need to find solutions $\bm\psi_\uparrow=(\rho_\uparrow,u_\uparrow,v_\uparrow)$ of the form in Eq.~\ref{eq:generalsolns}, under the conditions $u_\uparrow(y=0) = v_\uparrow(y=0) = 0$. 
Substituting in Eq.~\ref{eq:eigenvec}, this gives:
\begin{equation}\label{eq:NS-edge}
    \begin{pmatrix}
    0 \\ 0
    \end{pmatrix}
    =\left.\begin{pmatrix}
     u_\uparrow \\ v_\uparrow
    \end{pmatrix}\right|_{y=0}
    =\begin{pmatrix}
        U^+ & U^- \\
        i V^+ & i V^- 
    \end{pmatrix}
    \begin{pmatrix}
    A_\uparrow^+\\ A_\uparrow^-
    \end{pmatrix}
    =M\bm{A}_\uparrow
\end{equation}
Like in the previous section, a non-zero solution for $\bm A_\uparrow$ exists if and only if $\det M=0$, which in the gap only happens along $\omega=q$ as shown in Fig.~\ref{fig:sgndet}. 
Substituting this dispersion relation back into Eq.~\ref{eq:eigenvec} gives  $R^\pm = U^\pm\neq0$ and $V^\pm=0$. 
Eq.~\ref{eq:NS-edge} then tells us that $A_\uparrow^\pm \propto \pm 1/U_\pm$, giving $A_\pm\bm\Psi_\uparrow^\pm\propto\pm(1,1,0)^T$. 
Finally, substituting this into Eq.~\ref{eq:generalsolns} we find the no-slip edge solution $\bm\psi_\uparrow = (1,1,0)^T\sinh(\tfrac{1}{2}\Gamma y)e^{-\frac{1}{2}y}e^{i\omega(t-x)}$.
In terms of the $y$-profiles from the previous section, this one-sided waves on the top plane ($y>0$) are multiples of $K_1+K_1^a$. 
The one-sided waves on $y<0$ can than be found by sending $y \to -y$, giving $K_1-K_1^a$. 
Since these waves are completely decoupled from each other, the total space of admissible waves is $\mathrm{Span}\{K_1+K_1^a,K_1-K_1^a\}=\mathrm{Span}\{K_1,K_1^a\}$, in agreement with the previous section.
}

\subsection*{Appendix F: Fluid-fluid interface modes}
The stress-continuity condition on the fluid-fluid interface reads in vectorial form $[\bm v]=\bm0=\left[\bmh y\cdot\bm{\underline\sigma}\right]$. The latter of which can be expanded with
\begin{equation}
    \left(\bmh y\cdot\bm{\underline\sigma}\right)_j = 
    \delta_{2i}(-\rho\delta_{ij}-s\partial_i v^\perp_j)
    =
    (s\partial_y v , -s\partial_y u-\rho)_j .
\end{equation}
Again, substituting Eq.~(\ref{eq:generalsolns}) gives us a block matrix equation
\begin{equation}\label{eq:MFF}
    \begin{pmatrix}
    \bm 0 \\ \bm 0
    \end{pmatrix}
    =
    \begin{pmatrix}
    \underline{U} & -\underline{U} \\
    i\underline{V} & i\underline{V}
    \end{pmatrix}
    \begin{pmatrix}
    \bm A_\uparrow \\ \bm A_\downarrow
    \end{pmatrix}
    =
    \begin{pmatrix}
    \underline{U}(\bm A_\uparrow - \bm A_\downarrow) \\
    i\underline{V}(\bm A_\uparrow + \bm A_\downarrow)
    \end{pmatrix},
\end{equation}
with submatrices
\begin{equation}
    \underline{U} =
    \begin{pmatrix}
        U^+ & U^- \\
        \kappa_+ U^+ - R^+ & \kappa_- U^- - R^-
    \end{pmatrix}
\end{equation}
and
\begin{equation}
    \underline{V} =
    \begin{pmatrix}
        V^+ & V^- \\
        \kappa_+ V^+ & \kappa_- V^-
    \end{pmatrix}.
\end{equation}

This last matrix $\underline{V}$ is singular if and only if $\omega = q$. 
To see this, note that $\kappa_-$ never equals $\kappa_+$, since $\Gamma(\omega)>0$. 
Thus $\det\underline{V}=0$ requires either $V^+$ or $V^-$ to vanish, which happens if and only if $\omega\lambda_\pm = q\kappa_\pm$. 
Since $\lambda_\pm$ and $\kappa_\pm$ are both positive, we know that $\omega q\geq0$ and we may proceed to square both sides to obtain:
\begin{align}
    0 &= \omega^2(1\pm\Gamma(\omega))^2 - q^2(1\pm2\Gamma(\omega)+\Gamma^2(q))
    \nonumber\\
    &= (\omega^2-q^2)\left[(1\pm\Gamma(\omega))^2+4q^2\right].
\end{align}
The expression in square brackets is strictly nonzero, as $\Gamma(\omega)=1$ requires $|\omega|=\sqrt{m}>m$, which is outside the gap. 
In conclusion, either $\omega=q$ and $V^+=V^-=0$, or $\omega\neq q$ and $V^+$, $V^-$ are both non-zero. 
On the other hand, $\det\underline U$ vanishes along two curves that do not cross $\omega=q$, but are in the gap, as shown in Fig.~\ref{fig:sgndet}(b).

First, we again look at Kelvin waves along $\omega = q$. 
Here, $\underline V$ is the zero-matrix and $\underline U$ is invertible, thus by Eq.~(\ref{eq:MFF}), our solutions are even ($\bm A_\downarrow = \bm A_\uparrow$), but further unconstrained. 
This two-dimensional space of Kelvin waves can be spanned by functions $\bm\psi=(1,1,0)^TK(y)e^{-\frac{1}{2}|y|}e^{i\omega(t-x)}$ with hyperbolic functions in their $y$-profiles:
\begin{align}
    K_s = \sinh\left(\tfrac{1}{2}\Gamma|y|\right) 
    \quad\text{and}\quad
    K_c = \cosh\left(\tfrac{1}{2}\Gamma|y|\right).
\end{align}
We can orthogonalize this set as:
\begin{equation}
    K_1 = K_s 
    \quad\text{and}\quad
    K_2 = K_s - \Gamma K_c.
\end{equation}
The first is identical to the $K_1$ mode on the no-slip interface, while the second is orthogonal to both modes on the no-slip interface. 

In contrast to the no-slip interface, the fluid-fluid interface has boundary waves in the gap which do not lie along the line $\omega=q$. 
Since for these modes $\underline V$ is invertible, we see from Eq.~(\ref{eq:MFF}) that solutions must satisfy $\bm A_\downarrow = -\bm A_\uparrow$. 
Consequently, $\rho$ and $u$ are odd about $y=0$. 
Furthermore, $v$ is related to $\bm A$ with prefactor $is_\ud$ (see Eq.~(\ref{eq:eigenvec})), thus $v$ will be even and $90^\circ$ out of phase with both $u$ and $\rho$. 
This suggests a rotational flow for these so-called \emph{Yanai waves}, in contrast with the parallel flow of Kelvin waves.

These Yanai modes must satisfy $\underline{U}\bm A_\ud=\bm0$. 
Looking at the top row of $\underline U$, and knowing that $U^\pm\neq0$ from Eq.~(\ref{eq:eigenvec}), we see that $\bm A_\uparrow = -\bm A_\downarrow =(-U^-,+U^+)$. 
Thus along each line we find one non-degenerate mode $Y_1^o$ and $Y_2^o$, which we can write as
\begin{equation}\label{eq:generalsolns2}
    \bm\psi_\ud = s_\ud
    \left( U_+\Psi_\ud^+ e^{-\kappa_+|y|}
    -U_-\Psi_\ud^- e^{-\kappa_-|y|} \right)e^{i(\omega t-qx)}.
\end{equation}
The two modes are differentiated by the two Yanai dispersion curves $\omega(q)$ given by $\det\underline{U}=0$, for which we do not find a closed-form solution. 
Approximate values for this dispersion relation may be computationally obtained and plugged into Eqs.~(\ref{eq:kappa},\ref{eq:eigenvec}) to obtain numerical solutions.

\subsection*{Appendix G: Non-Hermitian Berry curvature}

\begin{figure}
    \centering
    \includegraphics[width=\linewidth, trim=0 -15 0 0]{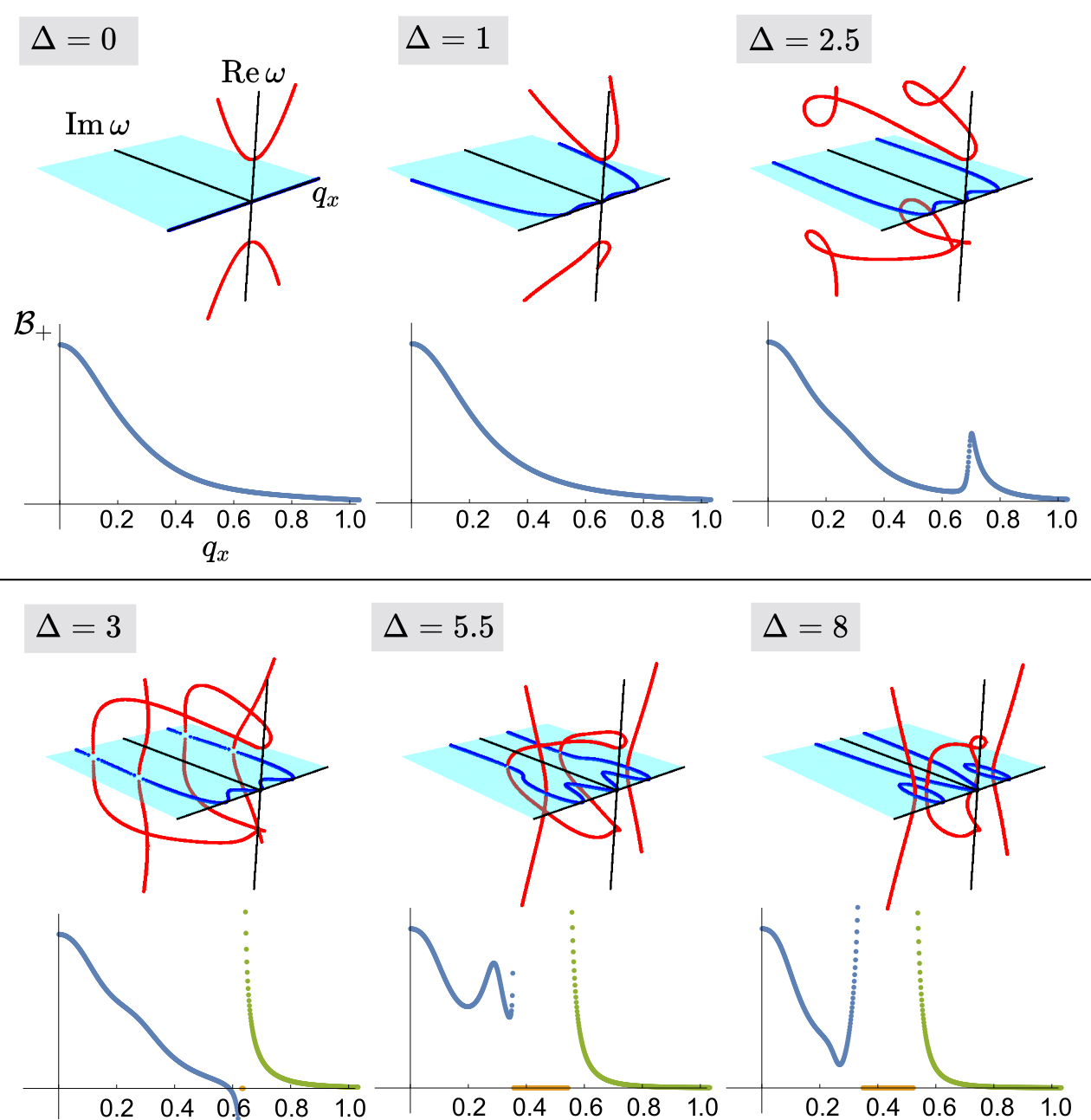}
    \caption{
    Plots of the band structure $\omega(\bm q)$ and Berry curvature of the top band $\mathcal B_+(\bm q)$, for different values of the dissipative difference parameter $\Delta \equiv D-\nu$. 
    We take a cut $\bm q =(q_x,0)$ in wavenumber space, which will look like any other cut due to radial symmetry. 
    As $\Delta$ increases, the imaginary parts of the eigenvalues become nonzero and develop geometric structure. 
    The middle band (blue) retains a zero real part, whereas the upper and lower bands (red) have reflection symmetry about the imaginary axis.
    While the viewing angle might suggest that red lines intersect, in 3D these lines do not touch.
    At $\Delta=2.5$, we see that the red bands have developed a secondary minimum in their real part which is pushed towards zero until the bottom and top bands meet the (blue) middle band, resulting in exceptional points.
    These exceptional points immediately separate as seen at $\Delta=3$. 
    Finally, the blue middle band pinches off to form rings in the imaginary plane, which are stable even beyond $\Delta=8$.
    In the bottom subfigures, we see a bump emerging in the Berry curvature, which diverges at the exceptional points. 
    In between the exceptional points the curvature is zero. 
    The colors for the curvature sections correspond to those in Fig.~\ref{fig:7}.
    These plots were calculated for $\nu=0$, which gives $\Delta =D$. 
    A non-zero $\nu$ adds a further imaginary component to each bands, which keeps the essential features (such as exceptional points) invariant.
    }\label{fig:NH}
\end{figure}

Our flux-trapping argument relies on protection from scattering, which typifies topological modes in Hermitian systems. 
This protection is guaranteed by a band gap in the bulk-spectrum, and the existence of these modes is provided by the non-zero Chern number along the top band. 
As one increases the value of the dissipative parameters, the non-Hermitian parts of the Hamiltonian will grow larger which will eventually close the band gap. 
These gap closings are themselves topological features called exceptional points, which form a cornerstone of current investigations into non-Hermitian topology.

To find the band-structure for the case of non-zero real viscosity $\nu$ and diffusion $D$ we need to solve the characteristic equation:

\begin{equation}\label{eq:NH-matrix}
    \begin{vmatrix}
        i \Delta q^2-\widetilde\omega & q_x & q_y \\
        q_x & -\widetilde\omega & \mp i (m-q^2) \\
        q_y & \pm i (m-q^2) & -\widetilde\omega
    \end{vmatrix} = 0,
\end{equation}
for $\omega(\bm q) = \widetilde\omega(\bm q) + i\nu q^2$, where $\Delta=D-\nu$ is the difference between the dissipative parameters. 
Note that the real parts of the bands $\omega(q)$ are invariant under any variation in dissipative parameters that keeps $\Delta$ constant, whereas the imaginary components of all bands differ by the same $\nu q^2$ term from the $\nu=0$ case. 
Since this will not change the existence of intersection points, we can plot the band structure for the parameters $\nu=0$ (but $D\ne0$) and use $m=1/5$ as we do in the main text.

In Fig.~\ref{fig:NH}, we plot these bands for various values of $\Delta$ along with the Berry curvature of the top band, calculated according to Eq.(\ref{eq:BerryCurv}). 
At $\Delta\approx2.8$ the top and bottom band meet at point along the middle band. 
This point splits immediately into two exceptional points which move apart along the middle band. 
For values of $q$ between these exceptional points, all three bands are completely imaginary, and no distinction can be made between the top, middle and bottom bands. 
This ambiguity is lifted by the fact that all eigenvectors with purely complex eigenvalues carry zero Berry curvature. 
This splits the Berry curvature for the top band in three parts, colored blue, yellow and green in Fig.~\ref{fig:NH}. 

Surprisingly, even though the Berry curvature diverges at the exceptional points, the total Berry curvature is well-behaved.
Numerical calculations shown in Fig.~\ref{fig:7} demonstrate that even for $\Delta > 2.8$ this total curvature remains finite. 
Surprisingly, we also note that the total Berry curvature is equally shared between the two separate sections when $\Delta > 2.8$.

\end{document}